\documentclass[lettersize,journal]{IEEEtran}
\IEEEoverridecommandlockouts

\usepackage{graphicx} 
\usepackage{braket}
\usepackage{subfigure}

\usepackage{xcolor}
\usepackage{threeparttable} 

\usepackage{tikz}
\definecolor{custompurple}{rgb}{0.5, 0.1, 0.5}
\usepackage{circuitikz}
\usetikzlibrary{quantikz}
\usepackage{setspace} 
\usepackage{circledsteps}
\usepackage{multirow}
\usepackage{caption}   

\pgfkeys{/csteps/inner ysep=10pt}
\usepackage{yquant}
\useyquantlanguage{groups}

\usepackage[normalem]{ulem}

\usepackage{amsmath,amsthm}

\usepackage{amssymb}

\usepackage{amsfonts}
\usepackage{comment}

\usepackage{algorithm}
\usepackage{algorithmicx}
\usepackage{algpseudocode}

\newtheorem{thm}{Theorem}

\newtheorem{defn}{\textbf{Definition}}
\newtheorem{exam}{Example}

\usepackage{enumitem}
\setlist[itemize,enumerate]{leftmargin=*}

\newcommand {\F} {{\mathcal{F}}}

\newcommand {\low} {{\texttt{low}}}
\newcommand {\high} {{\texttt{high}}}
\newcommand {\idx} {{\texttt{var}}}
\newcommand {\w} {{\texttt{op}}}

\begin{document}

\title{Oracle Synthesis Based on X-Map Decision Diagrams}


\author{
Xin Hong, Kezhen Zhang, Aochu Dai, Sanjiang Li, Shenggang Ying, Mingsheng Ying\\

\thanks{Work partially supported by Quantum Science and Technology-National Science and Technology Major Project under Grant No. 2024ZD0300502, Beijing Nova Program Grant No. 20240484652, and the National Natural Science Foundation of China Grant No. 12471437.
}

\thanks{
Xin Hong, Kezhen Zhang and Shenggang Ying are with Key Laboratory of System Software (Chinese Academy of Sciences), Institute of Software, Chinese Academy of Sciences, China. Aochu Dai is with the Department of Computer Science and Technology, Tsinghua University, China. Sanjiang Li and Mingsheng Ying are with the Centre for Quantum Software and Information, University of Technology Sydney, Australia.

}
}
        
\maketitle

\begin{abstract}

Quantum oracles act as reversible black-box operators that encode classical Boolean functions into quantum states, enabling efficient function evaluation in quantum superposition. The resource efficiency of oracle implementation is critical to the performance of numerous quantum algorithms. Most state-of-the-art oracle synthesis approaches rely on compact Boolean function representations such as exclusive-sum-of-products (ESOP), yet still suffer from excessive $T$-count and $CX$-count for large-scale functions.

In this paper, we propose a novel compact representation named X-Map decision diagram (XMDD) for Boolean functions, which integrates local invertible maps and complement edges to achieve higher compression efficiency. Based on XMDD, we further develop an optimized quantum oracle synthesis algorithm. Extensive experimental results demonstrate that, for Boolean functions with more than seven input variables, our method outperforms the state-of-the-art ESOP-based approach and Qiskit in nearly all test cases, achieving simultaneous reduction in both $T$-count and $CX$-count without an obvious trade-off. Moreover, we show that the performance can be further boosted by employing more optimal variable orderings. The proposed XMDD-based framework provides a scalable and resource-efficient solution for practical oracle synthesis in near-term and fault-tolerant quantum computing.


\end{abstract}


\section{Introduction}

Quantum computing has rapidly evolved from a theoretical model into a disruptive computational paradigm that promises exponential speed-ups for specific classes of problems. By exploiting characteristics such as superposition and entanglement, quantum algorithms, including Shor's factoring~\cite{shor1994algorithms}, Grover's search~\cite{grover1996fast}, and quantum machine learning kernels~\cite{schuld2019quantum}, outperform their best-known classical counterparts, thereby attracting substantial attention from both academia and industry.  Near-term Noisy Intermediate-Scale Quantum (NISQ) devices, however, impose stringent constraints on qubit count, gate fidelity, and coherence time; consequently, every primitive required by higher-level algorithms must be realized with minimal resources and depth.  

Among these primitives, \emph{quantum oracle}—the reversible embedding of a classical Boolean function—is of fundamental importance. Oracles are invoked repeatedly in Grover's algorithm, amplitude amplification, quantum walk, and many other routines, and their implementation cost frequently dominates the overall circuit depth, $T$-count, and ancilla budget. Thus, even modest reductions in oracle complexity can translate directly into larger problem instances that can be executed on current or near-future quantum hardware.

Over the past two decades, oracle synthesis has been intensively studied from multiple aspects. The earliest systematic approaches cast the problem as a two-step workflow: first derive a reversible Boolean network that realizes the given function, and second map that network into fault-tolerant Clifford+$T$ gates.  Exclusive-Sum-of-Products (ESOP) minimization falls into this paradigm: the function is written as an XOR of product terms, each product is turned into a multi-controlled $X$ (MCX) gate with appropriate controls, and the resulting set of $MCX$ gates is resynthesized for linear nearest-neighbour architectures or for $T$-depth reduction~\cite{fazel2007esop,saeedi2013synthesis,henderson2023automated,mishchenko2001fast}. These $MCX$ gates are further decomposed into $CX$ and $T$ gates via Barenco-style templates~\cite{barenco1995elementary} or Selinger’s $T$-par algorithm~\cite{selinger2013quantum}, and the cost can be reduced with the aid of ancilla qubits \cite{baker2019decomposing}. Reed–Muller (RM) expansions pursue a similar path but express the function as an XOR of conjunctions of fixed polarity; although canonical, RM forms are sensitive to variable ordering and tend to generate deep circuits for non-linear functions~\cite{green1986families}. Larger-scale functions are often tackled with hierarchical methods \cite{soeken2018lut,meuli2022xor,zhang2025divide}, where ancilla qubits are utilized to break down the functions into smaller parts that are easier to deal with. While conceptually simple, all of these techniques share a common weakness: the number of $T$ gates, $CX$ gates, and the ancilla count grow rapidly with the number of variables, often becoming the dominant cost in end-to-end quantum algorithms.

Our work is inspired by two recent results that successfully employ decision diagrams for quantum state preparation \cite{mozafari_efficient_2022, xin2025limtddqsp}.  Mozafari et al.~\cite{mozafari_efficient_2022} show that BDD-style structures can deterministically prepare quantum states with lower complexity than conventional approaches. After that, Vinkhuijzen et al. introduced local reversible maps into decision diagrams for the first time, resulting in the local invertible map decision diagram (LIMDD) model, which significantly enhances the efficiency of quantum state representation. Building on this, Hong et al. further propose the Local Invertible Map Tensor Decision Diagram (LimTDD) by incorporating local invertible maps into tensor representations \cite{hong_limtdd_2025}, and demonstrate competitive circuits for preparing arbitrary pure states~\cite{xin2025limtddqsp}. Apart from this, decision diagrams have been successfully applied to problems such as reversible logic synthesis \cite{wille2008synthesis}. These findings suggest that a similar diagrammatic strategy could be leveraged for the more specific but pivotal task of oracle synthesis, where the target is not a general state but the reversible embedding of a Boolean function.

Building on this insight, we propose an oracle synthesis algorithm that operates directly on the decision-diagram representation of Boolean functions. Motivated by the strong impact of diagram compactness on synthesis efficiency, we introduce a novel decision structure named the X-Map decision diagram (XMDD). By combining the strengths of LIMDD \cite{vinkhuijzen2023limdd} and complement-edged decision diagrams \cite{akers1978functional}, XMDD enables a significantly more compact encoding of Boolean functions.

By recursively processing diagram nodes and employing a single ancilla qubit to manage the activation of open and closed branches, our approach avoids costly multi-controlled gates along all computational paths, leading to simultaneous reductions in both $T$-count and $CX$-count. Extensive experimental results demonstrate that our method outperforms the state-of-the-art ESOP-based synthesis flow in nearly all test cases for Boolean functions with more than seven input variables, while introducing at most one additional ancilla qubit. Furthermore, the overall performance can be further improved by adopting more optimal variable orderings. These results confirm that XMDD provides a scalable and resource-efficient solution for practical oracle construction, advancing the realization of large-scale quantum algorithms on real quantum hardware.

The remainder of this paper is organized as follows. Section \ref{sec:background} reviews the necessary background on quantum computing, Boolean functions, and oracle synthesis. Section \ref{sec:Rep_LimTDD} introduces the XMDD framework and its application in representing Boolean functions. Section \ref{sec:basic_cons} presents key ingredients of our algorithm, including basic constructions and branch processing techniques, and Section \ref{sec:alg} presents the synthesis algorithm. Section \ref{sec:exp} provides a detailed example to illustrate the synthesis process. Section \ref{sec:complexity} analyses the time and gate complexity of the proposed method. Finally, Section \ref{sec:experiments} discusses experimental results, and Section \ref{sec:conclusion} concludes the paper.

\section{Background}\label{sec:background}

In this section, we first introduce some background on quantum computing, Boolean functions, and oracle synthesis.

\subsection{Quantum Computing}\label{subsec:qc}

\subsubsection*{1. Qubits}

Quantum computing is a computational paradigm that leverages the principles of quantum mechanics to process information in ways that are fundamentally different from classical computers. The basic unit of quantum information is the \emph{qubit} (quantum bit), which can exist in a superposition of the computational basis states $\ket{0}$ and $\ket{1}$, $$\ket{\psi} = \alpha\ket{0} + \beta\ket{1}, \quad \text{where } \alpha, \beta \in \mathbb{C} \text{ and } |\alpha|^2 + |\beta|^2 = 1.$$
This superposition allows quantum computers to represent and manipulate an exponentially large state space using only a polynomial number of qubits.

In this paper, we write $\ket{0}_{v}$ (or $\ket{0}_{n-1}$, $\ket{0}_{0}$, etc.) to mean $\ket{0}$ on the corresponding qubit.

\subsubsection*{2. Quantum Gates}

Quantum computations are performed by applying \emph{quantum gates}, which are unitary operators that evolve the state of the qubits. 
In this paper, we mainly care about the $T$, $X$, $CX$, and multi-controlled $X$ (MCX) gate.
Below we only introduce several fundamental gates that are widely used in oracle constructions.
\begin{itemize}
    \item \textbf{$T$ gate ($T$):} Introduces a phase:
    \[
    T\ket{0} = \ket{0}, \quad T\ket{1} = e^{i\pi/4}\ket{1}.
    \]
    In fault-tolerant quantum computing, achieving it usually requires a significant amount of resources.
    \item \textbf{Pauli-X gate (X):} Analogous to the classical NOT gate, it flips the basis states:
    \[
    X\ket{0} = \ket{1}, \quad X\ket{1} = \ket{0}.
    \]
    \item \textbf{Controlled-X gate (CX, CNOT):} Applies an X gate to the target qubit if the control qubit is in state $\ket{1}$:
    \[
    \text{CX}\ket{c}\ket{t} = \ket{c}\ket{t \oplus c}.
    \]
    The $CX$ gate is essential for creating entanglement and is a building block for multi-qubit controlled operations.
    \item \textbf{Controlled-$X$ with multiple conditions (C$^n$X):} Also known as the \emph{multi-controlled X} (MCX) gate, it applies an X gate to the target qubit if \emph{all} $n$ control qubits are in state $\ket{1}$. For example, the CCX (Toffoli) gate is a C$^2$X gate:
    \[
    \text{CCX}\ket{c_1}\ket{c_2}\ket{t} = \ket{c_1}\ket{c_2}\ket{t \oplus (c_1 \cdot c_2)}.
    \]
\end{itemize}
In this paper, we also employ quantum gates controlled by the state $\ket{0}$. We use the notation $\ket{p}$-controlled $X$ gates to indicate the cases when the activation condition for the $X$ gate is $p\in \{0,1\}^n$ for some $n$. In addition, we use the notation \(X_k\) to denote the single-qubit Pauli-X gate acting only on the $k$-th qubit. Formally, \(X_k\) corresponds to the multi-qubit unitary operator\(X_k = I \otimes \cdots \otimes I \otimes X \otimes I \otimes \cdots \otimes I,\)where the Pauli-$X$ gate appears at the $k$-th position and identity operators $I$ act on all other qubits. This definition ensures no ambiguity when \(X_k\) is multiplied or composed with other multi-qubit operators.

\subsubsection*{3. Quantum Circuits}

A \emph{quantum circuit} consists of a structured sequence of unitary quantum gates acting on a register of qubits, which are conventionally initialized to the ground state \(\ket{0}^{\otimes n}\) before any gate operation is executed. As the dominant computational model for modern quantum computing, the quantum circuit model is universal and can enable the approximation of any unitary evolution. Key structural properties, including the number of $T$ gates and multi-qubit gates contained in the circuit, directly affect the feasibility of execution on noisy intermediate-scale quantum (NISQ) devices as well as future fault-tolerant quantum platforms.

\subsection{Boolean Function}

A Boolean function over $n$ variables is a map $f:\{0,1\}^n\to\{0,1\}$ that assigns a binary output to every binary input string. The on-set of \(f\) is defined as $\text{On}(f) = \{ x \in \{0,1\}^n \mid f(x) = 1 \}$, and the off-set is defined to be $\text{Off}(f) = \{ x \in \{0,1\}^n \mid f(x) = 0 \}$.

Let $x_i$ be a variable of $f$; the positive and negative cofactors of $f$ with respect to $x_i$ are defined to be the Boolean functions
\[
f_{x_i}(x_1,\dots,x_{i-1},x_{i+1},\dots,x_n)= f(x_1,\dots,1,\dots,x_n),
\]
\[
f_{\overline{x}_i}(x_1,\dots,x_{i-1},x_{i+1},\dots,x_n)= f(x_1,\dots,0,\dots,x_n).
\]
Cofactors are the Boolean analogue of partial evaluation and are the basic blocks in Shannon decomposition $f= \overline{x_i}f_{\overline{x_i}}+x_if_{x_i}.$ We use \(\overline{x_i}\) to denote the logical negation (NOT) of the Boolean variable \(x_i\).

Let $f$ be a Boolean function among $n$ variables $x_{n-1},\cdots, x_0$, and $O = X^{b_{n-1}}\otimes \cdots \otimes X^{b_{0}}$ be a local operator of $n$ qubits, where each $b_i \in \{0,1\}$. We define the transformed function $O \odot f$ to be a Boolean function such that 
$$O \odot f = f[x_{n-1}/x_{n-1}',\cdots,x_{0}/x_{0}'],$$ 
where $x_i' = b_i\cdot \overline{x_i} + \overline{b_i}\cdot x_i$. In other words, the Pauli-X operators in $O$ flip the corresponding variables of $f$. In particular, this definition satisfies 
$\sum_{x\in On(O \odot f)}{\ket{x}} = O \sum_{x\in On(f)}{\ket{x}}$, where $\sum_{x\in On(f)}{\ket{x}}$ is the quantum state corresponding to $f$.

Note that this operation commutes with logical negation, that is, 
$$\neg (O \odot f) = O \odot \neg f.$$

\subsection{Oracle Synthesis}\label{subsec:os}

In many quantum algorithms, such as Grover's search, the target function is accessed through a quantum oracle—a black-box unitary operator \( U_f \) that encodes a classical Boolean function \( f: \{0,1\}^n \rightarrow \{0,1\} \). The oracle acts on the computational basis as:
\[
U_f |x\rangle|0\rangle = |x\rangle|f(x)\rangle,
\]
where \( x \in \{0,1\}^n \). This reversible embedding allows the function to be queried in superposition, a key requirement for quantum speedups. 

The challenge lies in efficiently implementing \( U_f \) using a quantum circuit. The cost of synthesis is typically measured in terms of gate count, especially $T$ count and $CX$-count. While existing approaches are effective in some cases, they often suffer from high gate complexity or require significant ancillary resources. This motivates the development of more efficient and scalable synthesis methods, such as the one proposed in this paper using XMDDs.

\section{X-Map Decision Diagrams}\label{sec:Rep_LimTDD}

Decision diagrams are the most commonly used way to compactly represent Boolean functions. Akers introduces complement edges for enhancing decision diagrams \cite{akers1978functional}. LIMDD \cite{vinkhuijzen2023limdd} and LimTDD \cite{hong_limtdd_2025} introduce local invertible maps into decision diagrams, demonstrating high compression efficiency in representing quantum states and tensors. XMDD combines these technologies to achieve a more compact encoding of Boolean functions. However, it should be noted that since XMDD is designed exclusively for Boolean functions, the local invertible maps on edges need only be composed of $I$ and $X$ operators and their tensor products.

\subsection{XMDD}\label{subsec:LimTDD}
Let $\mathcal{G}$ be the set of local operators of the form $X^{b_{n-1}}\otimes \cdots \otimes X^{b_{0}}$, where $n\in \mathbb{N}$ and each $b_i \in \{0,1\}$. 

\begin{defn}[XMDD]\label{def:limtdd}
 An XMDD $\mathcal{F}$ over a set of variables $S$ is a rooted, weighted, and directed acyclic graph 
	$\mathcal{F} = (V, E, \idx, \low, \high, \w, \texttt{neg})$ defined as follows:
	\begin{itemize}
		\item $V$ is a finite set of nodes which consists of non-terminal nodes $V_{NT}$ and a terminal node $v_T$ labelled with integer 1. Denote by $r_\mathcal{F}$ the unique root node of $\mathcal{F}$;
		\item $\idx: V_{NT} \rightarrow S$ assigns each non-terminal node an variable in $S$. We call $\idx(r_\mathcal{F})$ the top variable of $\F$, if $r_\mathcal{F}$ is not the terminal node;
		\item both $\low$ and $\high$ are mappings in $V_{NT} \rightarrow V$, which map each non-terminal node to its 0- and 1-successors, respectively;
		\item $E = \{(v, \low(v)), (v, \high(v)) : v\in V_{NT}\}$ is the set of edges, where $(v, \low(v))$ and $(v, \high(v)) $ are called the low- and high-edges of $v$, respectively. For simplicity, we also assume the root node $r_\mathcal{F}$ has a unique incoming edge, denoted  $e_r$, which has no source node;
		\item $\w: E\rightarrow \mathcal{G} \cup\{0, 1\}$ assigns to each edge a weight that is either a constant 0 or 1, or an operator in \(\mathcal{G}\).  $\w(e_r)$ is called the weight of $\mathcal{F}$, and denoted $w_\mathcal{F}$. 
		\item $\texttt{neg}: E\rightarrow \{0,1\}$  indicates whether a logical negation should be applied along an edge.        
	\end{itemize} 

\end{defn}

When representing Boolean functions, the semantics of the terminal node is defined to be the constant $1$, the semantics of an edge $e$, directing to a node $v$, is defined as
$$
f_e = \texttt{neg}(e)\oplus (\w(e) \odot f_v).
$$
In other words, if $\texttt{neg}(e)=0$, $f_e = \w(e) \odot f_v$; otherwise, $f_e = \neg(\w(e) \odot f_v)$. The semantics of a non-terminal node $v$ with $\idx(v)=x_n$ is defined to be 
$$
f_v = \overline{x_n}\cdot f_{(v,low(v))} + x_n\cdot f_{(v,high(v))}.\\
$$
The Boolean function $f_\F$ represented by an XMDD $\F$ is defined to be the Boolean function represented by the incoming edge $e_r$ of the root node $r_\F$. 

The introduction of local operators allows two Boolean functions that differ only by local transformations to share the same underlying structure, with the difference encoded in the edge operator alone. Similarly, the negation edge enables two mutually complementary functions to share the same representation. Together, these two mechanisms significantly reduce the number of nodes and paths in the decision diagram, directly lowering the complexity of the resulting oracle circuits. Furthermore, since negation commutes with the action of local operators, the two techniques can be safely combined without introducing ambiguity or conflicting semantics.

For clarity of presentation, we adopt a fixed variable order \(x_{n-1} \prec \cdots \prec x_0\) throughout the theoretical description. In the experimental section, however, we employ various variable orders to evaluate their influence and obtain optimal performance.

\begin{figure}[htbp]
    \centering
    \begin{minipage}{0.2\textwidth}
        \centering
        \resizebox{0.9\textwidth}{!}{




\begin{tikzpicture}[
>=latex,
line join=bevel,
every node/.style={minimum size=1.5cm, inner sep=0pt,thick, font=\fontsize{18pt}{22pt}}, 
every path/.style={line width=1.5pt} 
]
\node (1) at (128.67bp,18.0bp) [draw=red,circle,font=\fontsize{24pt}{28pt}] {$1$};
  \node (y01) at (46.668bp,108.9bp) [draw=red,circle,font=\fontsize{24pt}{28pt}] {$v_{00}$};
  \node (y02) at (185.67bp,108.9bp) [draw=red,circle,font=\fontsize{24pt}{28pt}] {$v_{01}$};
  \node (y10) at (66.668bp,202.96bp) [draw=red,circle,font=\fontsize{24pt}{28pt}] {$v_{10}$};
  \node (y11) at (165.67bp,202.96bp) [draw=red,circle,font=\fontsize{24pt}{28pt}] {$v_{11}$};
  \node (y2) at (107.67bp,297.02bp) [draw=red,circle,font=\fontsize{24pt}{28pt}] {$v_{20}$};
  \node (-0) at (107.67bp,387.93bp) [draw,draw=none] {};
  \draw [red,->,dotted] (y01) ..controls (29.551bp,79.314bp) and (25.128bp,64.774bp)  .. (32.668bp,54.0bp) .. controls (47.583bp,32.686bp) and (76.911bp,24.306bp)  .. (1);
  \definecolor{strokecol}{rgb}{0.0,0.0,0.0};
  \pgfsetstrokecolor{strokecol}
  \draw (53.668bp,61.875bp) node {};
  \draw [blue,->] (y01) ..controls (73.917bp,78.36bp) and (94.182bp,56.39bp)  .. (1);
  \draw (116.67bp,61.875bp) node {};
  \draw [red,->,dotted] (y02) ..controls (166.06bp,77.316bp) and (153.94bp,58.425bp)  .. (1);
  \draw (180.67bp,61.875bp) node {};
  \draw [blue,->] (y02) ..controls (205.77bp,80.723bp) and (212.16bp,65.699bp)  .. (204.67bp,54.0bp) .. controls (194.36bp,37.899bp) and (174.56bp,29.102bp)  .. (1);
  \draw (230.29bp,61.875bp) node {$0$};
  \draw [red,->,dotted] (y10) ..controls (31.234bp,190.19bp) and (11.932bp,180.12bp)  .. (2.6676bp,163.81bp) .. controls (-5.0407bp,150.24bp) and (6.2358bp,136.64bp)  .. (y01);
  \draw (23.668bp,155.93bp) node {};
  \draw [blue,->] (y10) ..controls (59.707bp,169.93bp) and (56.313bp,154.3bp)  .. (y01);
  \draw (80.543bp,155.93bp) node {$0$};
  \draw [red,->,dotted] (y11) ..controls (163.61bp,171.29bp) and (164.02bp,158.88bp)  .. (166.67bp,148.06bp) .. controls (167.5bp,144.65bp) and (168.68bp,141.19bp)  .. (y02);
  \draw (187.67bp,155.93bp) node {};
  \draw [blue,->] (y11) ..controls (193.01bp,184.7bp) and (203.51bp,175.28bp)  .. (208.67bp,163.81bp) .. controls (212.85bp,154.5bp) and (210.23bp,144.22bp)  .. (y02);
  \draw (230.42bp,155.93bp) node {$X$};
  \draw [red,->,dotted] (y2) ..controls (93.74bp,264.75bp) and (86.177bp,247.77bp)  .. (y10);
  \draw (110.67bp,249.99bp) node {};
  \draw [blue,->] (y2) ..controls (125.36bp,273.0bp) and (130.96bp,265.2bp)  .. (135.67bp,257.87bp) .. controls (140.87bp,249.76bp) and (146.13bp,240.69bp)  .. (y11);
  \draw (175.92bp,249.99bp) node {$X\otimes I$};
  \draw [red,->] (-0) ..controls (107.67bp,358.85bp) and (107.67bp,343.46bp)  .. (y2);
  \draw (158.54bp,344.05bp) node {$X\otimes I\otimes I$};
%
\node (neg) at (8.92bp,170.99bp) [fill,circle,minimum size=10pt] {};
\draw (105.543bp,60.93bp) node {$0$};
\draw (45.543bp,55.93bp) node {$0$};
\end{tikzpicture}

%
%
}
    \end{minipage}
    \hfill
    \begin{minipage}{0.25\textwidth}
        \small 
        \begin{align*}
            f_{\F_0} &=(X \otimes I \otimes I) \odot f_{v_{20}}\\
            f_{v_{20}} &= \overline{x_2} \cdot f_{v_{10}} + x_2 \cdot ((X \otimes I) \odot f_{v_{11}}) \\
            f_{v_{11}} &= \overline{x_1}\cdot f_{v_{01}} + x_1\cdot (X \odot f_{v_{01}}) \\
            f_{v_{10}}&=\overline{x_1}\cdot 1+x_1\cdot 0 = \overline{x_1}\\            
            f_{v_{00}}& =0\\
            f_{v_{01}}&=\overline{x_0}.
        \end{align*}
    \end{minipage}
    \caption{An XMDD representing the Boolean function $f(x) =x_2\overline{x_1}+\overline{x_2}x_1\overline{x_0}+\overline{x_2}\overline{x_1}x_0$. Black dots are used to denote that a logical negation is applied to the corresponding edge.
    }
    \label{fig:limtdd}
\end{figure}
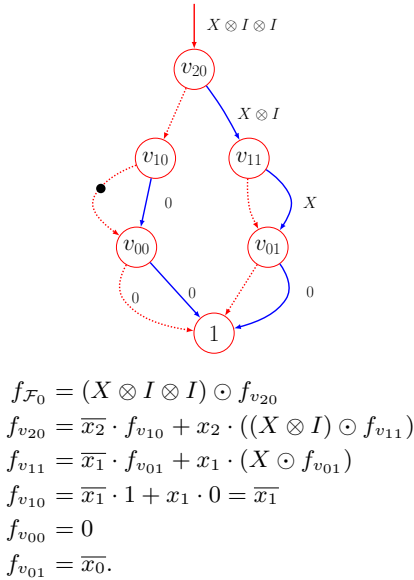

Following the conventions in \cite{vinkhuijzen2023limdd,hong_limtdd_2025}, we introduce the graphical notation for an XMDD node as
$\Circled{v_{0}}\overset{(n_0,O_0)}{\dashleftarrow}\Circled{v}\xrightarrow{(n_1, O_1)}\Circled{v_{1}}$. Here, \(n_0, n_1 \in \{0,1\}\) indicate whether a logical negation is applied to the corresponding edge, and \(O_0, O_1 \in \mathcal{G}\cup \{0\}\) denote the associated local operators. In figures such as Fig.~\ref{fig:limtdd}, we use a black dot to visually mark negation for simplicity.

When constructing the XMDD representation of an $n$-variable Boolean function $f$, we first compute the cardinality \(\#\mathrm{On}(f)\). If \(\#\mathrm{On}(f) > 2^{n-1}\) (i.e., \(\#\mathrm{On}(f) > \#\mathrm{Off}(f)\)), we construct the XMDD for \(\neg f\) instead and set the negation flag of the incoming edge to 1. The same strategy is applied recursively when building XMDDs for sub-functions. As a result, \(n_0\) and \(n_1\) will never be 1 simultaneously. We further apply a normalization step to standardize the operators on the two outgoing edges of each node, following the procedure in \cite{vinkhuijzen2023limdd, hong_limtdd_2025}. This ensures that \(O_0\) is always the identity operator $I$ unless it is 0. Certainly, it can also be 1 if the corresponding edge directs to the terminal node.

In addition, the whole diagram can be represented by the incoming edge:
$\xrightarrow{(n_{\mathcal{F}},w_{\mathcal{F}})}\Circled{r_\F}.$

\begin{exam}

Fig. \ref{fig:limtdd} illustrates the XMDD structure of the Boolean function $f(x) =x_2\overline{x_1}+\overline{x_2}x_1\overline{x_0}+\overline{x_2}\overline{x_1}x_0$. 

The terminal node corresponds to the constant Boolean function 1. An edge assigned weight 0 stands for the constant function 0, while other terminal-connected edges correspond to the constant function 1.

Accordingly, nodes \(v_{00}\) and \(v_{01}\) encode \(f_{v_{00}}=0\) and \(f_{v_{01}}=\overline{x_0}\), respectively.

Since the on-set of the low branch of \(v_{10}\) exceeds its off-set, we attach a negation to this edge, making it equivalent to the constant function 1. Meanwhile, the high edge of \(v_{10}\) represents the constant function 0. Consequently, node \(v_{10}\) encodes \(f_{v_{10}} = \overline{x_1}\). For node \(v_{11}\), its low edge inherits the function of \(v_{01}\), and its high edge corresponds to \(X\odot f_{v_{01}}\), which yields \(f_{v_{11}} = \overline{x_1}\overline{x_0} + x_1 x_0\).

By the same derivation rule, we have
\(f_{v_{20}} = \overline{x_2}\cdot f_{v_{10}} + x_2 \cdot \big((X\otimes I)\odot f_{v_{11}}\big)
= \overline{x_2}\overline{x_1} + x_2\left(x_1\overline{x_0}+\overline{x_1}x_0\right).\)The overall function encoded by the complete XMDD \(\mathcal{F}_0\) is
\(f_{\mathcal{F}_0} = (X\otimes I\otimes I)\odot f_{v_{20}}
= x_2\overline{x_1} + \overline{x_2}x_1\overline{x_0} + \overline{x_2}\overline{x_1}x_0,\)which is consistent with the target function.

\end{exam}

\section{Basic Constructions and Techniques} \label{sec:basic_cons}

We now show how to turn a Boolean function that is already represented as an XMDD into a quantum oracle circuit. 

In this paper, we refer to \(q_d\) as the data qubit, which is designated to store the output value of the Boolean function $f$. Then, \(X_d\) stands for the Pauli-$X$ gate acting only on this data qubit, which flips its stored binary state.

\subsection{Key Observations}

Our oracle-synthesis algorithm is based on three key observations that are stated as the following three theorems.

\begin{thm}\label{thm:th3}
    Let $f$ be a Boolean function. Suppose $U$ is an unitary operator such that $U\ket{x}\ket{0} = \ket{x}\ket{f(x)}$. Then $X_{d}\cdot U\ket{x}\ket{0} = \ket{x}\ket{\neg f(x)}$.
\end{thm}

\begin{thm}\label{thm:th1}
    Let $f$ be a Boolean function among $n$ variables $x_{n-1},$ $\cdots, x_0$, and $O=X^{b_{n-1}}\otimes \cdots \otimes X^{b_0}$, where each $b_i \in \{0,1\}$. Suppose $U$ is an unitary operator  such that $U\ket{x}\ket{0} = \ket{x}\ket{f(x)}$. Then $(O \otimes I)\cdot U\cdot (O\otimes I) \ket{x}\ket{0} = \ket{x}\ket{O\odot f(x)}$.
\end{thm}

\begin{thm}\label{thm:th2}
    Let $f$ be a Boolean function among $n+1$ variables $x_{n},\cdots, x_0$, and let $x = x_{n-1},\cdots, x_0$. Suppose $U_0$ and $U_1$ are two unitary operators such that $U_0\ket{x}\ket{0} = \ket{x}\ket{f_{\overline{x_n}}(x)}$ and $U_1\ket{x}\ket{0} = \ket{x}\ket{f_{x_n}(x)}$. Then $U_0 \oplus U_1 \ket{x_nx}\ket{0} = \ket{x_nx}\ket{f(x_nx)}$.
\end{thm}


\subsection{Basic Constructions}

The above three theorems lead us to introduce the following two Basic Constructions.

\noindent{\textbf{BC1 (Incoming Edge Operator Processing)}}\quad Let
    $\mathcal{F} = \xrightarrow{(n_{\mathcal{F}},O)} \Circled{v}$ be an XMDD, representing a Boolean function $f$. Suppose $U\ket{x}\ket{0} = \ket{x}\ket{f_{v}(x)}$. Then $X_d^{n_{\mathcal{F}}}\cdot (O \otimes I)\cdot U\cdot (O\otimes I) \ket{x}\ket{0} = \ket{x}\ket{f(x)}$.

\noindent{\textbf{BC2 (Recursive processing)}}\quad 
Let $\Circled{v_0} \overset{(n_0,I)}{\dashleftarrow} \Circled{v} \xrightarrow{(n_1,O)} \Circled{v_1}$ be a node in an XMDD with $\idx(v)=x_n$. Suppose $U_0\ket{x}\ket{0} = \ket{x}\ket{f_{v_0}(x)}$, $U_1\ket{x}\ket{0} = \ket{x}\ket{f_{v_1}(x)}$. Then, $(X_d^{n_0}\cdot U_0) \oplus (X_d^{n_1}\cdot(O\otimes I)U_1(O\otimes I))\ket{x_nx}\ket{0}=\ket{x_nx}\ket{f_v(x_nx)}$.

\begin{figure}[htbp]
\centering
\subfigure[]{
\begin{tikzpicture}
      \begin{yquant}[register/minimum height=5mm, operator/separation=2mm, control style={radius=1.7pt}, subcircuit box style={dashed}]
    
    qubit {$q_n$} q1;
    qubits {} q0;
    qubit {$q_d$} qd;
    
    slash q0;
    box {$X_d^{n_0}U_0$} (q0,qd) | ~q1;
    box {$O$} q0 | q1;
    box {$X_d^{n_1}U_1$} (q0,qd) | q1;
    box {$O$} q0 | q1;
  \end{yquant}
\end{tikzpicture}
}
\subfigure[]{
\begin{tikzpicture}
      \begin{yquant}[register/minimum height=5mm, operator/separation=2mm, control style={radius=1.7pt}, subcircuit box style={dashed}]
    
    qubit {$q_n$} q1;
    qubits {} q0;
    qubit {$q_d$} qd;
    
    slash q0;
    box {$X_d^{n_0}U_0$} (q0,qd) | ~q1;
    box {$O$} q0;
    box {$X_d^{n_1}U_1$} (q0,qd) | q1;
    box {$O$} q0;
  \end{yquant}
\end{tikzpicture}
}
\caption{(a) A circuit that implements $f_v$, where $v$ is a node $\Circled{v_0} \overset{(n_0,I)}{\dashleftarrow} \Circled{v} \xrightarrow{(n_1,O)} \Circled{v_1}$ with $\idx(v)=x_n$, $U_0$ and $U_1$ are circuits that implement $f_{v_0}$ and $f_{v_1}$, respectively; (b) An equivalent but simplified circuit.
}
\label{fig:synqc1}
\end{figure}
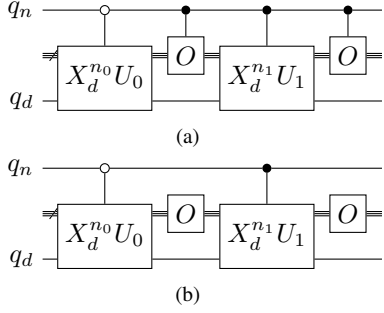

The operator $(X_d^{n_0}\cdot U_0) \oplus (X_d^{n_1}\cdot(O\otimes I)U_1(O\otimes I))$ can be implemented by a circuit shown in Fig. \ref{fig:synqc1} (a). Specifically, we notice that the two $\ket{1}$-controlled $O$ gates can be replaced by two $O$ gates directly as shown in Fig. \ref{fig:synqc1} (b).

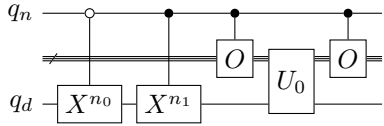
\begin{figure}[htbp]
\centering

\begin{tikzpicture}
      \begin{yquant}[register/minimum height=5mm, operator/separation=2mm, control style={radius=1.7pt}, subcircuit box style={dashed}]
    
    qubit {$q_n$} q1;
    qubits {} q0;
    qubit {$q_d$} qd;
    
    slash q0;
    box {$X^{n_0}$} qd | ~q1;
    box {$X^{n_1}$} qd | q1;
    box {$O$} q0 | q1;
    box {$U_0$} (q0,qd);
    box {$O$} q0 | q1;
  \end{yquant}
\end{tikzpicture}

\caption{
A circuit that implements $f_v$, where $v$ is a node $\Circled{v_0} \overset{(n_0,I)}{\dashleftarrow} \Circled{v} \xrightarrow{(n_1,O)} \Circled{v_0}$ with $\idx(v)=x_n$ and $U_0$ is a circuit that implements $f_{v_0}$.
}
\label{fig:synqc2}
\end{figure}

We also notice that, when $U_0=U_1$, the circuit shown in Fig.~\ref{fig:synqc1} can be reduced to the circuit shown in Fig.~\ref{fig:synqc2}.

The above BC1 copes with the weights on the incoming edge, while BC2 enables us to transform a problem of oracle synthesis with $n$ variables into two problems of oracle synthesis with $n-1$ variables. By repeating this process until reaching a trivial situation, we can generate a circuit that represents this oracle. 
However, one drawback of doing it this way is that we would need to use an extremely large number of control qubits for the quantum gates.

\begin{figure}[htbp]
\centering
\begin{tikzpicture}
      \begin{yquant}[register/minimum height=5mm, operator/separation=2mm, control style={radius=1.7pt}, subcircuit box style={dashed}]
    qubit {$q_n$} q1;
    qubit {$q_{n-1}$} q1p;
    qubits {} q0;
    qubit {$q_d$} qd;
    slash q0;
    box {$U_{00}$} (q0,qd) | ~q1, q1p;
    box {$U_{01}$} (q0,qd) | q1p, ~q1;
    box {$U_{10}$} (q0,qd) | q1, ~q1p;
    box {$U_{11}$} (q0,qd) | q1,q1p;
  \end{yquant}
\end{tikzpicture}
\caption{The synthesis framework obtained by iteratively using BC2. We omit the processing of operators and negations on edges.}
\label{fig:bc2_r}
\end{figure}
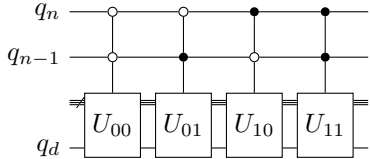

Note that we adopt the decomposition scheme illustrated in Fig.~\ref{fig:synqc1} whenever a branch node is encountered, namely a node with distinct zero and one successors (\(v_0\neq v_1\)). This scheme introduces an additional control qubit for constructing recursive subcircuits. By contrast, the structure in Fig.~\ref{fig:synqc2} is applicable only to non-branch nodes satisfying \(v_0=v_1\), which effectively reduces such overhead.

Accordingly, during node processing, the total number of required control qubits is determined by the count of branch nodes along the path from the root to the target node, while the corresponding control constraints are specified by the assignments of these branching nodes, termed branch conditions.
For example, consider an XMDD whose top two layers are fully expanded. The circuit structure obtained by iteratively applying construction rule BC2 (Fig. \ref{fig:synqc1}) twice is presented in Fig. \ref{fig:bc2_r}. As illustrated, when implementing the component \(U_{ij}\) corresponding to the $ij$-branch (i.e., the $j$-th successor of the $i$-th successor of the root node), we adopt the branch condition \(\ket{i}_n\ket{j}_{n-1}\) as the control state. For concise illustration, we omit the quantum gates responsible for coping with edge operators and negation in this diagram.

\subsection{An Ancilla Qubit Can Help}

This issue can be mitigated by introducing an ancilla qubit \( q_a \), similar to the technique used in \cite{mozafari_efficient_2022,xin2025limtddqsp}. This ancilla qubit marks certain branches (nodes or paths) of the decision diagram as "open" while others are marked as "closed". Here, "open" signifies that a branch is active and can be processed, whereas "closed" indicates that a branch is inactive and cannot be processed, with the control condition \(|1\rangle_a\). When dealing with a branch \( p \), we make sure that \( q_a \) is in state \(|1\rangle_a\) only if the most significant qubits are in \(|p\rangle\). This means that branch $p$ is marked open by $q_a$, allowing operators initially controlled by \(|p\rangle\), which can contain many qubits, to now be controlled solely by \(|1\rangle_a\).

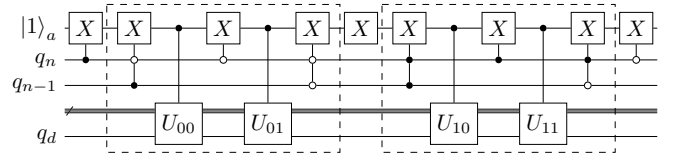
\begin{figure}[htbp]
\centering
\resizebox{0.48\textwidth}{!}{
\begin{tikzpicture}
      \begin{yquant}[register/minimum height=3mm, operator/separation=0.7mm, control style={radius=1.5pt}, subcircuit box style={dashed}]
      
    qubit {$\ket{1}_a$} qa;
    qubit {$q_n$} q1;
    qubit {$q_{n-1}$} q1p;
    qubits {} q0;
    qubit {$q_d$} qd;
    slash q0;
    x qa | q1;
    subcircuit {
      qubit {} qa;
      qubit {} q1;
      qubit {} q1p;
      qubits {} q0;
      qubit {} qd;
    x qa | q1p, ~q1;
    box {$U_{00}$} (q0,qd) | qa;
    x qa | ~q1;
    box {$U_{01}$} (q0,qd) | qa;
    x qa | ~q1p, q1;
    } (qa,q1,q1p,q0,qd);
    x qa;
    subcircuit {
      qubit {} qa;
      qubit {} q1;
      qubit {} q1p;
      qubits {} q0;
      qubit {} qd;
    x qa | q1p, q1;
    box {$U_{10}$} (q0,qd) | qa;
    x qa | q1;
    box {$U_{11}$} (q0,qd) | qa;
    x qa | q1,~q1p;
    } (qa,q1,q1p,q0,qd);
    
    x qa | ~q1;
  \end{yquant}
\end{tikzpicture}}

\caption{The synthesis framework obtained by iteratively using BC2 and introducing an ancilla qubit. We also omit the processing for operators and negations on edges.}
\label{fig:bc2_r2}
\end{figure}

The specific implementation method is as follows:

\textbf{Initialization:} Set \( q_a \) to \(|1\rangle_a \), marking the root node and entire XMDD as open.

\textbf{Recursion:} For a node \(\Circled{v_0} \overset{(n_0,I)}{\dashleftarrow} \Circled{v} \xrightarrow{(n_1,O)} \Circled{v_1}\) with branch condition \( p \), marked open by \( q_a \):
\begin{itemize}
    \item To process its 0-successor \( v_0 \), add a \(|p\rangle|1\rangle_v\)-controlled \( X \) gate to \( q_a \). This closes the high-branch of \( v \) and leaves only the low branch open.
    \item Using \(|1\rangle_a \) as the control condition, process the low branch.
    \item Apply a \(|p\rangle\)-controlled \( X \) gate to \( q_a \) to toggle the open/closed state of the branches.
    \item Use \(|1\rangle_a \) again to control the processing of the high branch.
    \item Finally, apply a \(|p\rangle|0\rangle_v\)-controlled \( X \) gate to \( q_a \) to reopen \( v_0 \), restoring node \( v \) to an open state and resetting \( q_a \).
\end{itemize}

The synthesis framework after applying this process iteratively to the second level is shown in Fig.~\ref{fig:bc2_r2}. It can be seen that when we deal with $U_{ij}$ part, \( q_a \) is in \(|1\rangle_a \) if and only if \( q_n \) and \( q_{n-1} \) are in \(|i\rangle_n|j\rangle_{n-1} \). Thus, the previous requirement of using \( (|i\rangle_n|j\rangle_{n-1} ) \) as the control condition is now simplified to using only \( \ket{1}_a \).

It is important to note that this process is activated only when $v_0 \neq v_1$. Otherwise, node \( v \) is not a branch node, and we can directly use $\ket{1}_a$ as the control condition to handle the two branches of the $v$ node.

Furthermore, this optimization does not always yield performance gains. When handling nodes in the lower half layers of the decision diagram, the extra control qubits needed to regulate ancilla qubits exceed those required by direct recursive construction. For this reason, we only adopt this strategy for the upper half layers of the diagram.

\section{Algorithm}\label{sec:alg}

The pseudocode of our algorithm is described in  Alg. \ref{alg:OSyn_onea}. Before entering the algorithm, we first cope with the operator and negation on the incoming edge of the root node. Then, we enter the recursive process.

When the two successors of node $v$ are identical, we could cope with its successor directly. If the high edge weight of $v$ is 0, a control condition $\ket{0}_v$ should be added. Otherwise, we use the circuit structure as shown in Fig. \ref{fig:synqc2}.

When the two successors of $v$ are distinct, and $v$ is a node at the upper half of the decision diagram, we first add a $\ket{p}\ket{1}_v$-controlled $X$ gate to $q_a$, closing the high-branch. Then we cope with 0-successor as well as the weights on the low edge. Here, $p$ is the branch condition of $v$. After that, we apply a $\ket{p}$-controlled $X$ gate to open the high-branch and close the low-branch. Then, we cope with the high-branch, where two $O$ operators should be added before and after the circuit block. Finally, we add a $\ket{p}\ket{0}_v$-controlled $X$ to reopen the low branch and make the node $v$ open again. For nodes located in the lower half layers, we directly adopt the construction scheme shown in Fig. \ref{fig:synqc1} (b).

\begin{algorithm}[htbp]
\caption{$\textsc{O\_Syn}(v, q_a, q_d, p)$}
\begin{algorithmic}[1]
\Require{A node $v$ of an XMDD $\F$ representing an $(n-|p|)$-variable Boolean function $f_v$; an ancilla qubit $q_a$ which marked $v$ open under the $|p|$-bit branch condition $p$. For the root node, the branch condition is empty.}
\Ensure{A quantum circuit $C$ with unitary $U$ such that $U \ket{1}_a\ket{p}\ket{x}\ket{0} = \ket{1}_a\ket{p}\ket{x}\ket{f_v(x)}$.}

\State $C \gets \mathrm{QuantumCircuit}(n+2)$ {\color{gray}\Comment{Empty quantum circuit}}
\If{$v$ is the terminal node}
\State Append to $C$ an $\ket{1}_a$-controlled $X$ operator on $q_d$
\State \Return $C$
\EndIf
\State Let $\w\big((v,\high(v))\big)=O$. 
\If{$\low(v)=\high(v)$}
\State $C_0 \gets \textsc{O\_Syn}\big(\low(v),q_a,q_d,p\big)$

\State Add to $C$ a $\ket{0}_v$-controlled $X$ gate on $q_d$, if $\texttt{neg}((v,\low(v))) = 1$
\State Add to $C$ a $\ket{1}_v$-controlled $X$ gate on $q_d$, if $\texttt{neg}((v,\high(v))) = 1$

\If{$O = 0$}
\State Append $C$ with $\ket{0}_v$-controlled $C_0$ if $\w{(v,\low(v))} \neq 0$
\Else
\State Add to $C$ a $\ket{1}_v$-controlled $O$ gate
\State Append $C$ with $C_0$ if $\w{(v,\low(v))} \neq 0$
\State Add to $C$ a $\ket{1}_v$-controlled $O$ gate
\EndIf
\Else
\State $C_0 \gets \textsc{O\_Syn}\big(\low(v),q_a,q_d,p0\big)$
\State $C_1 \gets \textsc{O\_Syn}\big(\high(v),q_a,q_d,p1\big)$
\If{$v$ is on the upper half of the original XMDD}
\State Add to $C$ a $\ket{p}\ket{1}_v$-controlled $X$ gate on $q_a$ to close the high-branch of $v$
\State Add to $C$ a $\ket{1}_a$-controlled $X$ gate on $q_d$, if $\texttt{neg}((v,\low(v))) = 1$
\State Append $C$ with $\ket{1}_a$-controlled $C_0$
\State Add to $C$ a $\ket{p}$-controlled $X$ gate on $q_a$ to close the low-branch and open the high-branch of $v$
\State Add to $C$ a $O$ gate
\State Add to $C$ a $\ket{1}_a$-controlled $X$ gate on $q_d$, if $\texttt{neg}((v,\high(v))) = 1$
\State Append $C$ with $\ket{1}_a$-controlled $C_1$
\State Add to $C$ a $O$ gate
\State Add to $C$ a $\ket{p}\ket{0}_v$-controlled $X$ gate on $q_a$ to open the low-branch of $v$
\Else
\State Add to $C$ a $\ket{1}_a\ket{0}_v$-controlled $X$ gate on $q_d$, if $\texttt{neg}((v,\low(v))) = 1$
\State Append $C$ with $\ket{1}_a\ket{0}_v$-controlled $C_0$
\State Add to $C$ a $O$ gate
\State Add to $C$ a $\ket{1}_a\ket{1}_v$-controlled $X$ gate on $q_d$, if $\texttt{neg}((v,\high(v))) = 1$
\State Append $C$ with $\ket{1}_a\ket{1}_v$-controlled $C_1$
\State Add to $C$ a $O$ gate
\EndIf
\EndIf
\State \Return $C$
\end{algorithmic}
\label{alg:OSyn_onea} 
\end{algorithm}

\section{An Example}\label{sec:exp}

\begin{figure}[htbp]
\centering

\resizebox{0.46\textwidth}{!}{
\begin{tikzpicture}
  \begin{yquant}[register/minimum height=1.3mm, operator/separation=1.2mm, control style={radius=2pt}, subcircuit box style={dashed}]
    qubit {$\ket{1}_a$} a;
    qubit {$q_2$} q2;
    qubit {$q_1$} q1;
    qubit {$q_0$} q0;
    qubit {$\ket{0}_d$} qd;
    x q2;
    x a | q2;
    
    subcircuit {
      qubit {} qa;
      qubit {} q2;
      qubit {} q1;
      qubit {} q0;
      qubit {} qd;
      x qd | qa,~q1;
    } (a,q2,q1,q0,qd);
    
    x a;
    
    subcircuit {
      qubit {} qa;
      qubit {} q2;
      qubit {} q1;
      qubit {} q0;
      qubit {} qd;
      x q1;
      x q0 | q1;
      x qd | qa,~q0;
      x q0 | q1;
      x q1;
      } (a,q2,q1,q0,qd);
      
    x a | ~q2;
    
    x q2;
  \end{yquant}
\end{tikzpicture}
}
\caption{The quantum synthesis circuit generated by Alg.~\ref{alg:OSyn_onea} for the Boolean function encoded by the XMDD \(\mathcal{F}_0\) in Fig.~\ref{fig:limtdd}. 
}
\label{fig:pre_cir}
\end{figure}
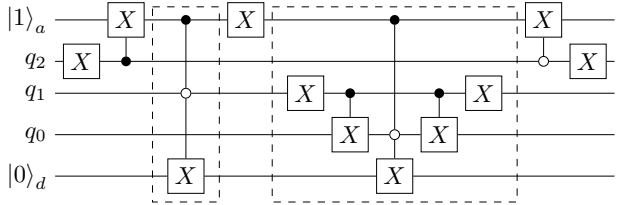

In this section, we illustrate the proposed oracle synthesis algorithm step-by-step using the Boolean function represented by the XMDD shown in Fig.~\ref{fig:limtdd}. The resulting circuit is shown in Fig.~\ref{fig:pre_cir}. 

An ancilla qubit \( q_a \), initialized to \( |1\rangle \), is used, marking the entire diagram as "open". The data qubit $q_d$ is initially set to be $\ket{0}_d$. Then, the initial state can be represented as $\ket{1}_a\ket{x_2x_1x_0}\ket{0}_d$. The Step-by-Step reduction is displayed as follows.

\subsection{Step-by-Step Reduction}

\subsubsection{Incoming edge operator}

We first cope with the weight $X\otimes I\otimes I$ on the incoming edge. An $X$ gate is added to $q_2$, making the state: $\ket{1}_a\ket{\overline{x_2}x_1x_0}\ket{0}_d$.

\subsubsection{Close 1-branch}

Then we use a $\ket{1}_2$-controlled X gate to close the 1-branch of the $v_{20}$ node. The state becomes: $\ket{x_2}_a\ket{\overline{x_2}x_1x_0}\ket{0}_d$.

\subsubsection{Processing the 0-branch}

Now, we turn to process the 0-branch. Note that there is no further branch node in this branch, but there is a negation on the low edge of $v_{10}$. So a $\ket{1}_a\ket{0}_1$-controlled X gate is applied, making the state: $\ket{x_2}_a\ket{\overline{x_2}x_1x_0}\ket{(x_2\cdot\overline{x_1})}_d$. Since the operator on the high edge of $v_{10}$ and the two outgoing edges of $v_{00}$ are all 0, no further gates need to be applied.

\subsubsection{Flip $q_a$}

Then we add an $X$ gate to adjust the open/close of the two successors of $v_{20}$, the state becomes: $\ket{\overline{x_2}}_a\ket{\overline{x_2}x_1x_0}\ket{(x_2\cdot\overline{x_1})}_d$.

\subsubsection{Processing the 1-branch}

\begin{itemize}
    \item Apply an $X$ gate to cope with the high-edge operator of $v_{20}$, making the state: $\ket{\overline{x_2}}_a\ket{\overline{x_2}\overline{x_1}x_0}\ket{(x_2\cdot\overline{x_1})}_d$.
    \item Add a $\ket{1}_1$-controlled $X$ gate to cope with high-edge operator of $v_{11}$, making the state: $\ket{\overline{x_2}}_a\ket{\overline{x_2}\overline{x_1}(x_0\oplus \overline{x_1})}\ket{(x_2\cdot\overline{x_1})}_d$.
    \item Add a $\ket{1}_a\ket{0}_0$-controlled $X$ gate to process the $v_{01}$ node, making the state: $\ket{\overline{x_2}}_a\ket{\overline{x_2}\overline{x_1}(x_0\oplus \overline{x_1})}\ket{(x_2\cdot\overline{x_1})\oplus (\overline{x_2}\cdot (\overline{x_0\oplus\overline{x_1}}))}_d=\ket{\overline{x_2}}_a\ket{\overline{x_2}\overline{x_1}(x_0\oplus \overline{x_1})}\ket{(x_2\cdot \overline{x_1}+\overline{x_2}\cdot x_1\overline{x_0}+\overline{x_2}\cdot \overline{x_1}\cdot x_0)}_d$.
    \item Add a $\ket{1}_1$-controlled $X$ gate to recover the high-edge operator of $v_{11}$, making the state: $\ket{\overline{x_2}}_a\ket{\overline{x_2}\overline{x_1}x_0}\ket{(x_2\cdot \overline{x_1}+\overline{x_2}\cdot x_1\overline{x_0}+\overline{x_2}\cdot \overline{x_1}\cdot x_0)}_d$.
    \item Add a $X$ gate to recover the high-edge operator of $v_{20}$, making the state: $\ket{\overline{x_2}}_a\ket{\overline{x_2}x_1x_0}\ket{(x_2\cdot \overline{x_1}+\overline{x_2}\cdot x_1\overline{x_0}+\overline{x_2}\cdot \overline{x_1}\cdot x_0)}_d$.
\end{itemize}

\subsubsection{Reopen 0-branch}

We use a $\ket{0}_2$-controlled X gate to reopen the 0-branch of the $v_{20}$ node. The state becomes: $\ket{1}_a\ket{\overline{x_2}x_1x_0}$ $\ket{(x_2\cdot \overline{x_1}+\overline{x_2}\cdot x_1\overline{x_0}+\overline{x_2}\cdot \overline{x_1}\cdot x_0)}_d$.

\subsubsection{Recover the operator on the high-edge of $v_{20}$}

Finally, we add an $X$ to recover the operator on the high-edge of $v_{20}$, making the state $\ket{1}_a\ket{x_2x_1x_0} \ket{(x_2\cdot \overline{x_1}+\overline{x_2}\cdot x_1\overline{x_0}+\overline{x_2}\cdot \overline{x_1}\cdot x_0)}_d $, thus, finishing the oracle synthesis process.

\section{Complexity}\label{sec:complexity}

We derive simple worst-case complexity bounds for the algorithm presented in Section \ref{sec:alg}.

In the following analysis, n denotes the total number of qubits, and p stands for the number of reduced paths within the XMDD. In our synthesis framework, nodes with identical child successors are processed only once. Accordingly, we merge their two outgoing edges when estimating computational complexity, and such merged edge routes are defined as reduced paths. For instance, the XMDD shown in Fig. \ref{fig:limtdd} contains merely two reduced paths.

\subsection{Time Complexity}

In our algorithm, we traverse the decision diagram in a depth-first manner, with each reduced path being traversed exactly once. Note that there are at most $n$ non-terminal nodes on a reduced path, each requiring handling of the high-edge operator and negation, and adjusting the open and close state using the ancilla qubit. Also, notice that there are at most $n$ local operators for any operator on the edges. The time complexity of our algorithm is $\mathcal{O}(n^2p)$.

\subsection{Gate Complexity}

Similarly, the complexity of the number of gates also depends on the number of reduced paths.

We first consider the operators appearing on the edges of the XMDD. There are no more than $n+(n-1)+\cdots+2+1 = \frac{n(n+1)}{2} $ local operators on each path, and these operators will be processed through $X$ or $CX$ gates. 

In addition to this, we need one $MCX$ gate to cope with the branch itself, with no more than $n+1$ control qubits and at most $n$ $MCX$ gates for coping with negations.

Then, we examine the quantum gates used to control the opening and closing of branches. For each branch node, we need one controlled gate to close its 1-branch, one controlled gate to flip the open/close status of both branches, and one controlled gate to reopen the 0-branch. Assuming there are $k$ branch nodes preceding the current node on the path, this requires two $(k+1)$-qubit controlled gates and one $k$-qubit controlled gate. 

In summary, the upper bound of gate requirements is:
\begin{itemize}
    \item $\mathcal{O}(np)$ $MCX$ gates, each with 2 to $n+1$ control qubits.
    \item $\mathcal{O}(n^2p)$ $CX$ gates.
    \item $\mathcal{O}(n^2p)$ X gates.
\end{itemize}

\section{Experiments}\label{sec:experiments}


We performed a comprehensive comparison between our method, the state-of-the-art ESOP-based method implemented in \cite{riener_easy_2024}, and the widely-used Qiskit tool \cite{javadi2024quantum}. Both our method and Qiskit are implemented in Python and can be directly executed in standard Python environments. In contrast, the ESOP-based approach is implemented in C++ and requires manual compilation via CMake before execution. All experiments are executed on a laptop equipped with 64GB of memory and an i9-13900H processor.


\subsection{Comparison on $T$ gates}

We represent each Boolean function as an XMDD, then use our algorithm for oracle synthesis, and count the number of $T$ gates and $CX$ gates required to implement the resulting circuit within the Clifford+$T$ gate set. We compare this count with the results obtained from the ESOP method and Qiskit. Note that the total number of $n$-variable Boolean functions equals \(2^{2^n}\), which exhibits extremely rapid exponential growth. This number reaches 65,536 when there are 4 variables and exceeds 429 billion with 5 variables. For this reason, we only sample 20,000 random Boolean functions for experimental evaluation when the number of input variables exceeds four. When the number of variables is small $(n=2, 3)$, we test all the Boolean functions.


\begin{figure}
    \centering
    \includegraphics[width=\linewidth]{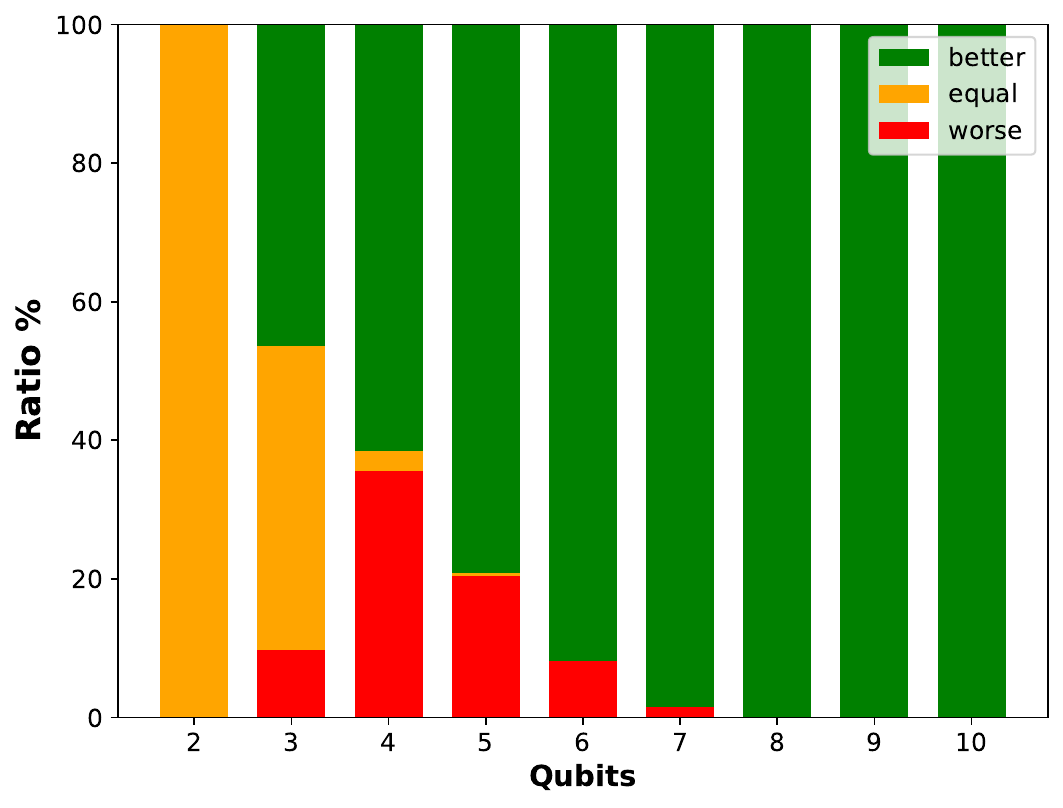}
    \caption{The comparison of our method and the ESOP method on the number of $T$ gates.
    }
    \label{fig:T_ESOP_3bar}
\end{figure}

\begin{figure}
    \centering
    \includegraphics[width=\linewidth]{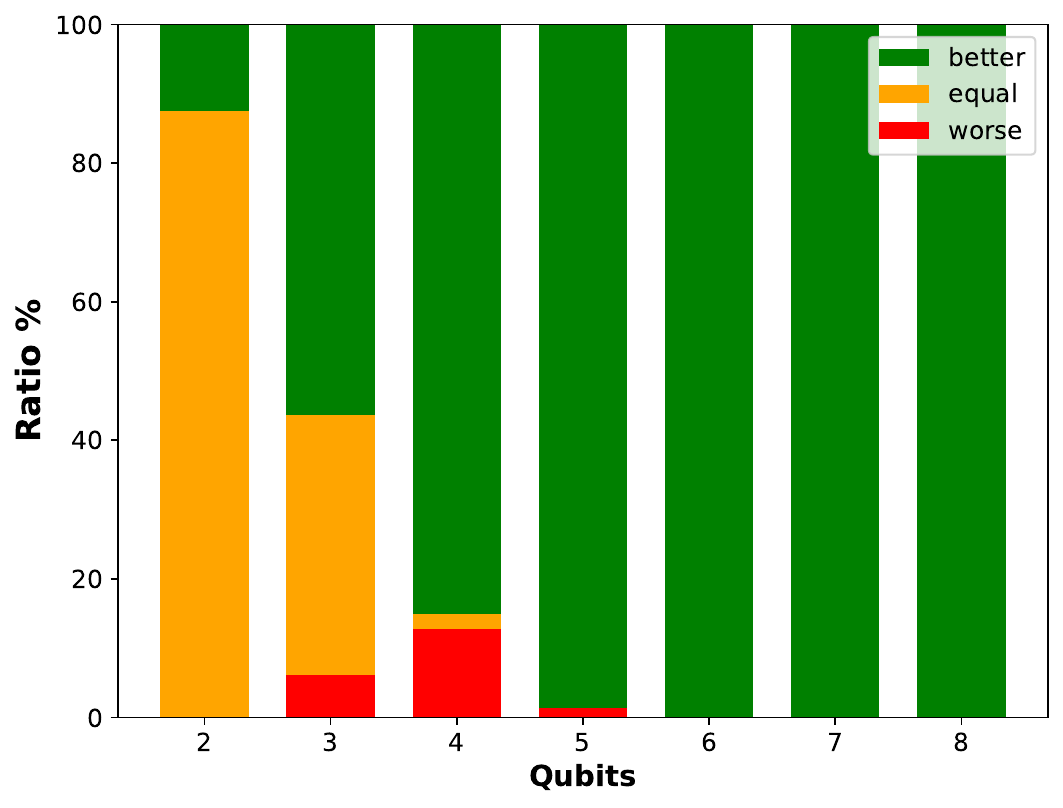}
    \caption{The comparison of our method and Qiskit on the number of $T$ gates.
    }
    \label{fig:T_Qis_3bar}
\end{figure}

\begin{table}[]
\caption{Experimental data for $T$ gates}
\resizebox{0.45\textwidth}{!}{
\begin{tabular}{llllllllll}
\hline
\multicolumn{1}{c}{\multirow{2}{*}{n}} & \multicolumn{1}{c}{\multirow{2}{*}{\#cases}} &  & \multicolumn{3}{c}{ESOP}    &  & \multicolumn{3}{c}{Qiskit} \\ \cline{4-6} \cline{8-10} 
\multicolumn{1}{c}{}                   & \multicolumn{1}{c}{}                         &  & lower   & same    & higher  &  & lower   & same   & higher  \\ \hline
2                                      & 16                                          &  & 0\% & 100\% & 0\%  &  &12.5\%         &87.5\%        &0\%         \\
3                                      & 256                                          &  & 46.5\% & 43.8\% & 9.8\%  &  &56.2\%         &37.5\%        &6.2\%         \\
4                                      & 20000                                        &  & 61.5\% & 3.0\% & 35.5\% &  &85.1\%         &2.1\%        &12.8\%         \\
5                                      & 20000                                        &  & 79.2\% & 0.4\%  & 20.4\% &  &98.7\%         & 0.0\%       & 1.3\%        \\
6                                      &20000                                          &  &91.8\% &0.1\%  &8.1\%   &  &100.0\%        &0\%        &0.0\%         \\ 
7                                      &20000                                          &  &98.5\% &0.0\%  &1.5\%   &  &100\%         &0\%        &0\%        \\ 
8                                      &20000                                          &  &100\%  &0\%    &0\%    &  &100\%         &0\%        &0\%         \\ 
9                                      &20000                                          &  &100\%  &0\%    &0\%    &  &-         &-        &-         \\ 
10                                     &20000                                          &  &100\%  &0\%    &0\%      &  &-         &-        &-         \\ 
\hline
\end{tabular}}
\begin{tablenotes}
\small
\item Note: In this table, 100\% and 0\% denote exact values, while 100.0\% and 0.0\% represent values that have been rounded.
\end{tablenotes}
\label{Tab:exp_data_t}
\end{table}


Figs. \ref{fig:T_ESOP_3bar} and \ref{fig:T_Qis_3bar} present a comparative analysis of the number of $T$ gates required by the oracle circuits synthesized by our XMDD-based method, the ESOP-based approach, and the Qiskit framework, respectively. The results are grouped by the number of input variables $n$, and each stacked bar is partitioned into three categories: better (green), equal (orange), and worse (red), which denote the proportion of test cases where our method yields fewer, equal, or more $T$ gates than the corresponding baseline. The detailed ratios are also provided in Table \ref{Tab:exp_data_t}.

For the comparison with the ESOP method (Fig. \ref{fig:T_ESOP_3bar}), our approach already achieves superior results in 61.5\% of cases at \(n=4\), and the proportion of better cases grows rapidly with the problem size, reaching 98.5\% at \(n=7\) and 100\% for \(n \ge 8\). Meanwhile, the share of worse cases drops significantly, from 35.5\% at \(n=4\) to 0\% for \(n \ge 8\). Compared with Qiskit (Fig. \ref{fig:T_Qis_3bar}), our method demonstrates even stronger performance: it outperforms Qiskit in 85.1\% of cases at \(n=4\), and 100.0\% of cases for \(n \ge 6\).


\subsection{Comparison on $CX$ gates}

\begin{figure}
    \centering
    \includegraphics[width=\linewidth]{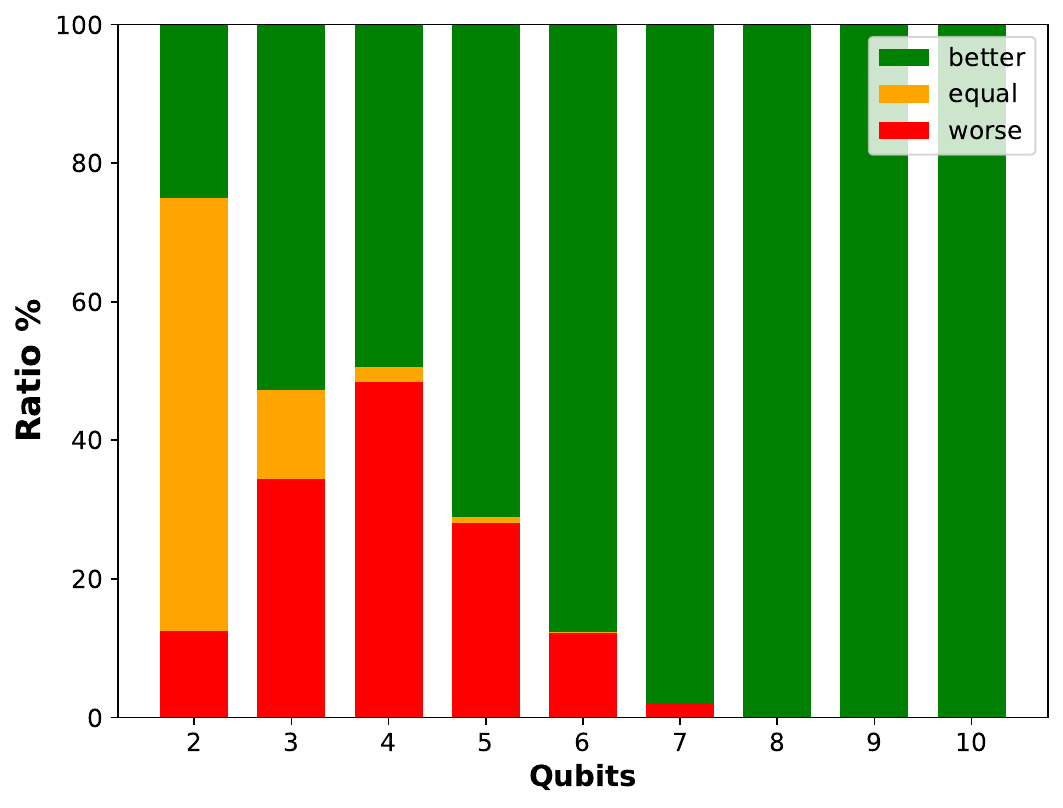}
    \caption{The comparison of our method and the ESOP method on the number of $CX$ gates.
    }
    \label{fig:CX_ESOP_3bar}
\end{figure}

\begin{figure}
    \centering
    \includegraphics[width=\linewidth]{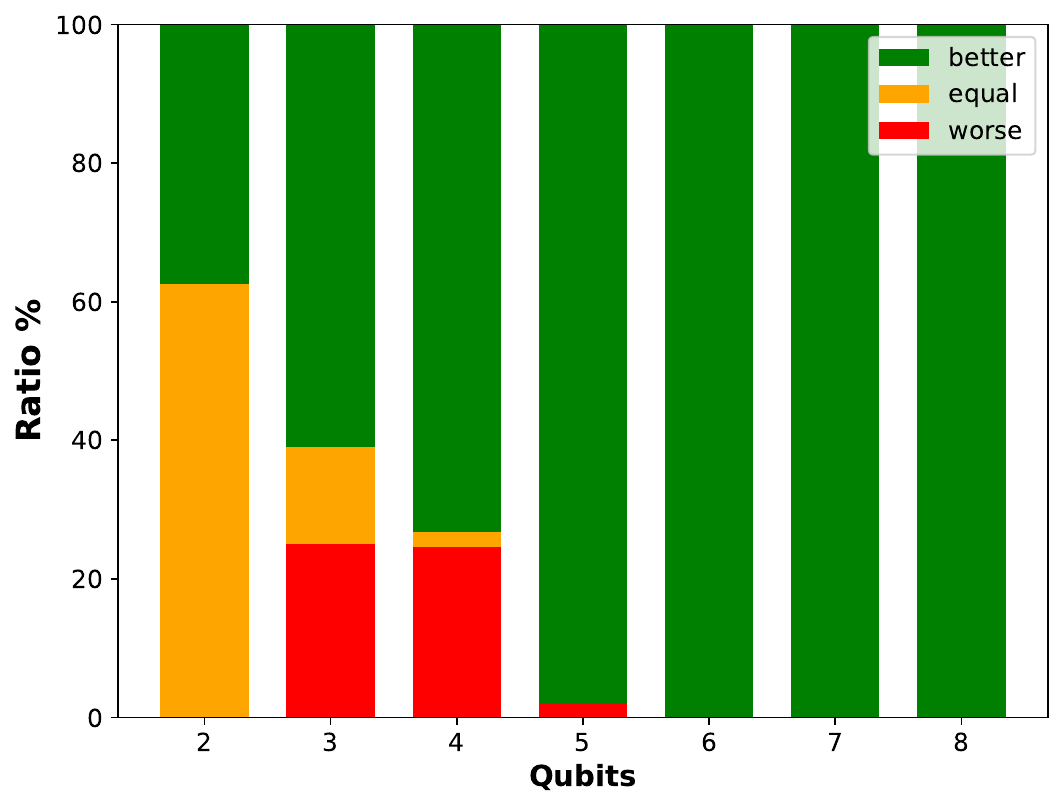}
    \caption{The comparison of our method and Qiskit on the number of $CX$ gates. 
    }
    \label{fig:CX_Qis_3bar}
\end{figure}

\begin{table}[]
\caption{Experimental data for $CX$ gates}
\resizebox{0.45\textwidth}{!}{
\begin{tabular}{llllllllll}
\hline
\multicolumn{1}{c}{\multirow{2}{*}{n}} & \multicolumn{1}{c}{\multirow{2}{*}{\#cases}} &  & \multicolumn{3}{c}{ESOP}    &  & \multicolumn{3}{c}{Qiskit} \\ \cline{4-6} \cline{8-10} 
\multicolumn{1}{c}{}                   & \multicolumn{1}{c}{}                         &  & lower   & same    & higher  &  & lower   & same   & higher  \\ \hline
2                                      & 16                                          &  & 25\% & 62.5\% & 12.5\%  &  &37.5\%         &62.5\%        &0\%         \\
3                                      & 256                                          &  & 52.7\% & 12.9\% & 34.4\%  &  &60.9\%         &14.1\%      &25.0\%         \\
4                                      & 20000                                        &  & 49.3\% & 2.2\% & 48.4\% &  &73.2\%         &2.1\%        &24.6\%         \\
5                                      & 20000                                        &  & 71.0\% & 1.0\%  & 28.0\% &  &98.0\%         & 0.1\%       & 1.9\%        \\
6                                      &20000                                          &  &87.6\% &0.2\%  &12.2\%   &  &100.0\%        &0\%        &0.0\%         \\ 
7                                      &20000                                          &  &98.0\% &0.0\%  &1.9\%   &  &100\%         &0\%        &0\%        \\ 
8                                      &20000                                          &  &100\%  &0\%    &0\%    &  &100\%         &0\%        &0\%         \\ 
9                                      &20000                                          &  &100\%  &0\%    &0\%    &  &-         &-        &-         \\ 
10                                     &20000                                          &  &100\%  &0\%    &0\%      &  &-         &-        &-         \\ 
\hline
\end{tabular}}
\begin{tablenotes}
\small
\item Note: In this table, 100\% and 0\% denote exact values, while 100.0\% and 0.0\% represent values that have been rounded.
\end{tablenotes}
\label{Tab:exp_data_CX}
\end{table}

Figs. \ref{fig:CX_ESOP_3bar} and \ref{fig:CX_Qis_3bar} present the comparison of $CX$ gate counts between our XMDD-based oracle synthesis method and the ESOP-based and Qiskit baselines, respectively. The Boolean function used here is exactly the same as that in the above subsection. As with the $T$-count evaluation, each stacked bar categorizes results into three groups: better (green, fewer $CX$ gates), equal (orange), and worse (red, more $CX$ gates). The detailed ratios are summarized in Table \ref{Tab:exp_data_CX}.

Compared with the ESOP method (Fig. \ref{fig:CX_ESOP_3bar}), our approach already outperforms in 52.7\% of cases at \(n=3\), with this ratio growing to 98.0\% at \(n=7\) and reaching 100\% for \(n\ge 8\). Although a larger proportion of cases initially require more $CX$ gates (e.g., 48.4\% at \(n=4\)), this disadvantage diminishes rapidly with increasing problem size, dropping to below 2\% for \(n\ge 7\). Against Qiskit (Fig. 10), our method exhibits even stronger advantages: it reduces $CX$ counts in 60.9\% of cases at \(n=3\), and achieves 100\% better performance for functions with \(n\ge 6\), with nearly no cases requiring more gates. 

Notably, our method does not involve a trade-off between $T$-count and $CX$-count reduction. In contrast, our XMDD-based method consistently achieves simultaneous improvements in both metrics across most test cases. The fundamental reason for this is that our method achieves a reduction in the number of $MCX$ gates through the high compression efficiency of XMDD. This joint reduction in both gate types is particularly significant for practical quantum implementation, as it directly lowers the overall circuit depth and error rate, making the resulting oracle circuits more suitable for execution on near-term quantum hardware.

\subsection{Detailed Comparison}

\begin{table}[]
\caption{Experimental data}
\resizebox{0.49\textwidth}{!}{
\begin{tabular}{llllllllll}
\hline
\multicolumn{1}{c}{\multirow{2}{*}{n}} & \multicolumn{1}{c}{\multirow{2}{*}{\#cases}} &  & \multicolumn{3}{c}{Average $T$ gate}    &  & \multicolumn{3}{c}{Average $CX$ gate} \\ \cline{4-6} \cline{8-10} 
\multicolumn{1}{c}{}                   & \multicolumn{1}{c}{}                         &  & XMDD   & ESOP    & Qiskit  &  & XMDD   & ESOP   & Qiskit  \\ \hline
2                                      & 16                                          &  & 3.5 &3.5  & 5.3  &  &3.6         &3.8        &5.0         \\
3                                      & 256                                          &  & 14.3 & 18.1 & 25.0  &  &14.6         &16.3      &22.0         \\
4                                      & 20000                                        &  & 52.3 & 58.5 & 78.2 &  &50.3         &50.8        &66.5         \\
5                                      & 20000                                        &  & 137.8 & 179.0  & 307.7 &  &126.7         &152.5       &258.3        \\
6                                      &20000                                          &  &400.0 &552.5  &974.7   &  &358.8        &466.7        &823.0         \\ 
7                                      &20000                                          &  &1081 &1657  &3708   &  &936         &1400        &3171       \\ 
8                                      &20000                                          &  &2492  &4943    &10973    &  &2142         &4196        &9390         \\ 
9                                      &20000                                          &  &7250  &14327    &-    &  &6196         &12206        &-         \\ 
10                                     &20000                                          &  &16184  &38579    &-      &  &13791         &32929        &-         \\ 
\hline
\end{tabular}}
\begin{tablenotes}
\small
\item Note: For simplicity, we retain one decimal place for results with \(n\le 6\), and round all values to integers for \(n\ge 7\).
\end{tablenotes}
\label{Tab:exp_data}
\end{table}


\begin{figure*}
    \centering
    \includegraphics[width=0.9\linewidth]{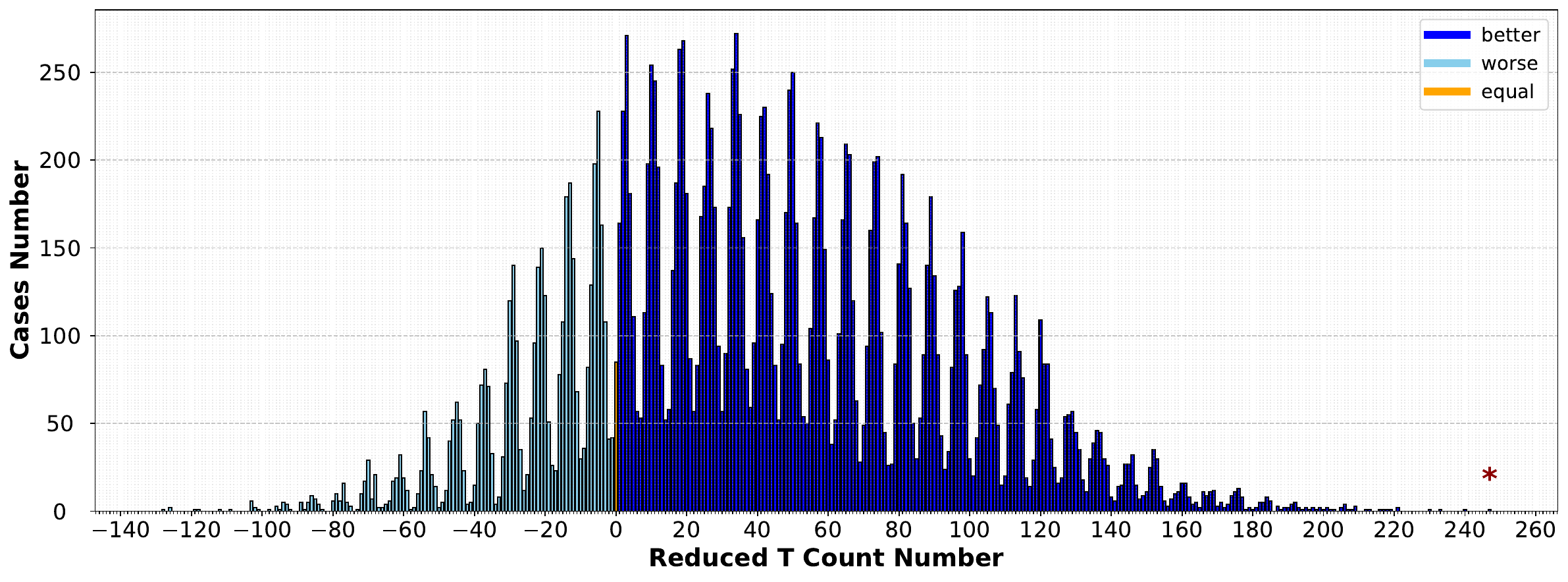}
    \caption{The distribution of the reduction in the number of $T$ gates for 20,000 five-variable Boolean functions. 
    }
    \label{fig:var5_T}
\end{figure*}

\begin{figure*}
    \centering
    \includegraphics[width=0.9\linewidth]{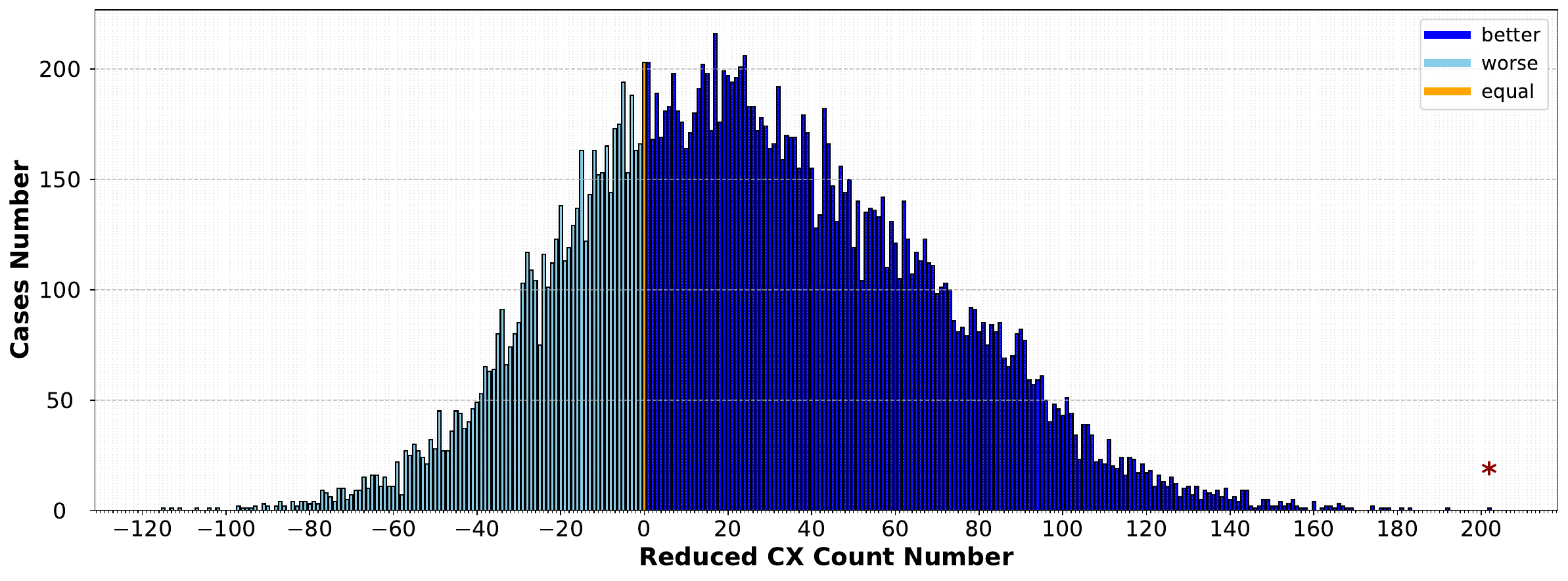}
    \caption{The distribution of the reduction in the number of $CX$ gates for 20,000 five-variable Boolean functions. 
    }
    \label{fig:var5_CX}
\end{figure*}

Figs. \ref{fig:var5_T} and \ref{fig:var5_CX} illustrate the distribution of $T$-count and $CX$-count reductions achieved by our XMDD-based method relative to the ESOP baseline, across 20,000 randomly generated 5-variable Boolean functions. Since Qiskit's overall performance is inferior to that of ESOP, we omit the comparison of our method with Qiskit in this section. The horizontal axis represents the number of reduced gates (positive values indicate our method uses fewer gates, negative values indicate more gates), while the vertical axis denotes the number of test cases. In both distributions, the majority of cases lie on the positive side of the axis, forming a right-skewed peak, which clearly shows that our method reduces both $T$ and $CX$ gates in most instances. The corresponding detailed average gate counts for different variable sizes are summarized in Table \ref{Tab:exp_data}.

As shown in Table \ref{Tab:exp_data}, the average $T$-count and $CX$-count of our method are consistently lower than both the ESOP-based and Qiskit-based baselines for all \(n \ge 4\). For example, for \(n=5\), our method reduces the average $T$ gates from 179.0 (ESOP) and 307.7 (Qiskit) to 137.8, and the average $CX$ gates from 152.5 (ESOP) and 258.3 (Qiskit) to 126.7. As the number of variables increases, the performance gap widens significantly: for \(n=8\), our method achieves only 2492 average $T$ gates and 2142 average $CX$ gates, while the ESOP baseline requires 4943 $T$ gates and 4196 $CX$ gates, and Qiskit requires as many as 10973 $T$ gates and 9390 $CX$ gates. These results confirm that our method delivers substantial and consistent reductions in both $T$-count and $CX$-count, with advantages becoming more pronounced for larger Boolean functions.

It is worth noting that we do not include runtime comparison here due to implementation in different programming languages: ESOP is implemented in C++, while our approach and Qiskit are developed in Python. As C++ generally exhibits significantly higher execution efficiency than Python, such a direct runtime comparison would be unfair. Nevertheless, our experimental results demonstrate that our algorithm consistently outperforms Qiskit in overall speed. While it is slower than ESOP for small-scale functions, our approach achieves comparable or even higher efficiency than ESOP in many cases when the number of variables is 9 or greater.


In addition, we examined all possible variable orders for small-scale Boolean functions to investigate the impact of variable order on the performance of our algorithm. Notably, adopting the optimal variable order can further reduce the resulting gate costs. With optimal variable ordering, our method achieves fewer or equal $T$ gates in 99.7\% of test cases for $n=3$, 91.4\% for $n=4$, 98.5\% for $n=5$, 99.6\% for $n=6$, and 100\% for $n=7$. The same optimization effect also applies to the reduction of $CX$ gate counts.




\section{Conclusion}\label{sec:conclusion}

In this paper, we present a novel and efficient quantum oracle synthesis framework based on the proposed X-Map decision diagram (XMDD). By combining local invertible maps and complement edges, XMDD enables a highly compact representation of Boolean functions, laying a solid foundation for low-resource oracle construction. The proposed synthesis algorithm recursively traverses the XMDD structure and uses a single ancilla qubit to manage branch activation, avoiding expensive multi-controlled operations and achieving simultaneous reduction in $T$-count and $CX$-count.

Experimental evaluations over a wide range of Boolean functions show that our method consistently surpasses the state-of-the-art ESOP-based synthesis and the Qiskit compiler, especially for functions with a large number of input variables. The advantage becomes increasingly significant as the variable count grows, and can be further enhanced with more optimal variable orderings.

These results validate that the XMDD-based oracle synthesis is scalable, resource-efficient, and robust, making it highly suitable for constructing lightweight oracles for NISQ devices and future fault-tolerant quantum systems.

\section*{Acknowledgment}
In preparing this paper, we used the generative AI tool Doubao to conduct grammar checking and polishing of the entire manuscript.

\bibliographystyle{IEEEtran}
\bibliography{reference}

\end{document}